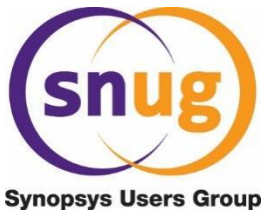

# In-design DFM rule scoring and fixing method using ICV

Vikas Tripathi, Yongfu Li, Zhao Chuan Lee,
I-Lun Tseng, Jason Khaw and Jonathan Ong

Globalfoundries Singapore Pte. Ltd.
Singapore
www.globalfoundries.com

**ABSTRACT**

*As compared to DRC rules, DFM rules are a list of selected recommended rules which aim to improve the design margins for better manufacturability. In GLOBALFOUNDRIES, we use DFM scoring methodology as an effective technique to analyze design quality in terms of manufacturability. Physical design engineers can perform our Manufacturability Check Deck (MCD) to asset their design quality during the sign-off stage. In the past, Synopsys users have to convert their design though milkyway database to GDSII format and execute the verification through the third party EDA tools. This method is costly and time-consuming for our Synopsys users. Today, we propose a new and easy-to-use integrated flow which leverages on the ICV engine to provide DFM scoring and in-design fixing techniques. The new methodology address DFM violations early in the design flow and achieve DFM compliance design during sign-off phase.*

# Contents

# Table of Figures

# 1. Introduction

Design For Manufacturability (DFM) rules are related to design rules with higher margin requirements. In GLOBALFOUNDRIES, we use to provide these rules as a part of Manufacturability Check Deck (MCD). MCD verify the design for all the DFM rule violation and produce an equivalent score. These scores are used to analyze design quality in terms of manufacturability and uses as DFM sign-off for tape-out. Designs with higher scores are considered as good quality designs. Such designs are also good candidates for potential high yield and less likely to have manufacturability or process related issues during fabrication.

Fig. 1 Illustrate an example of enclosure rule for three cases A, B and C and the difference between DRC and DFM results. DRC results can be PASS or FAIL. If the geometries satisfy the DRC requirements then it is DRC PASS otherwise it is DRC fail. The approach in DFM scoring checks is different. All the DFM violation is provided a score. If violation is at DRC margin or below DRC margin then DFM SCORE value will be "0" and if the drawn margin meeting DFM requirement then score is "1". Between DRC margin and DFM margin score will be between 0-1.

Here case A is drawn with value following DRC margin while case B is drawn with margin between DRC to DFM value and case C is drawn at DFM margin. During DRC checks, all three cases are passing as all fulfilling DRC requirements. But in DFM check all three cases result in different scores. Case A has given a score of "0" since it is drawn t DRC margin while case B has score in between 0-1 as margin are in between DRC and DFM and case C has full score of 1 as it full fill the DFM margin.

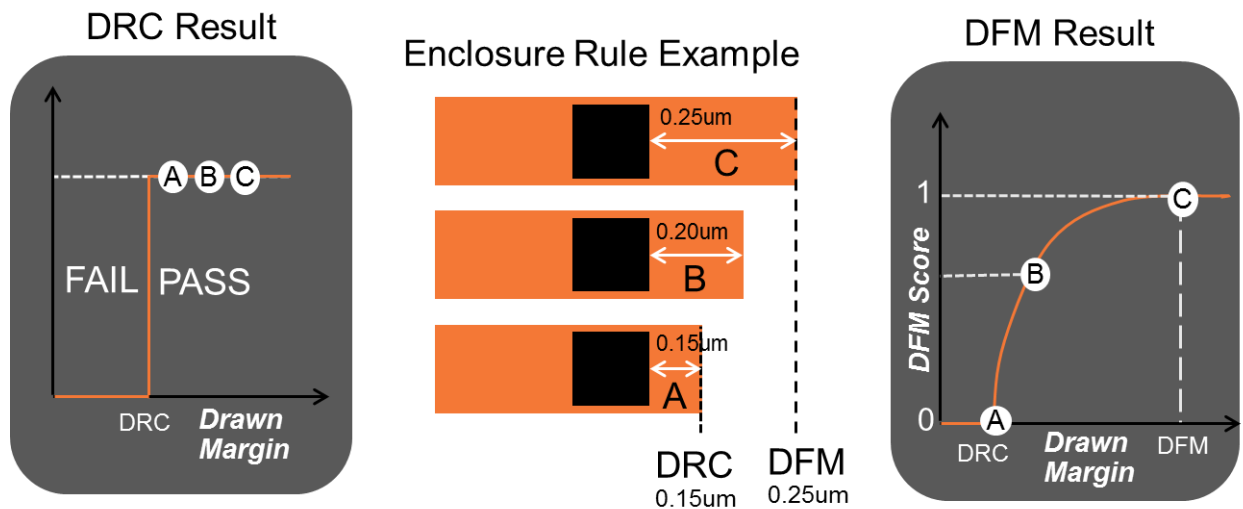

**Figure 1:** Showing difference in DRC and DFM check.

All violations with different scores combined together to produce total DFM SCORE value. This total score represents overall design compliance to DFM requirement. The total score compute based on yield scoring equations which we will discuss later part of this paper.

Foundries characterize these rules based on test chips or silicon data to get various parameter such as DFM margin, failure rate and process sensitivity etc., using silicon learning approach and by drawing multiple test chips. These test chips includes various test structures also known as DOE "Design of Experiment". These structures designed to capture various process effects like printability, etch etc.

## 2. Challenges

With the shrink in feature sizes the number of manufacturability issues increase [1]. This increases the number of physical verification checks and number of DRC & DFM rules. Implementation of DFM rules requires more margins and if applied through router then it may over-constrain the router and can impact higher computation time. Such approach may also end up with larger design area penalty. DFM scoring method can be very useful to avoid such scenario. This method helps to evaluate the design quality based on current DFM rule violation exists in the design. Since this method depends on aggregated scoring method so does not required to fix all the DFM violations and still design full fill DFM compliance requirement.

For ICC users, generally designers stream out their designs after completion of routing and other physical verifications and use third party EDA tools to run such DFM scoring verification. This approach will not only create dependency on third party EDA tools but also very inefficient as it require multiple iterations to close the DFM rule scoring and fixing loop. Such approach can also be very complicated and creates other challenges which required optimizing design to make it DFM compliance. This creates a requirement of in-design DFM rule scoring and fixing flow.

## 3. In-design approach of running DFM rule scoring and fixing technique

### 3.1 Flow overview

Fig 2 shows DFM scoring flow integrated in design cycle. DFM scoring checks should be triggered on placement and routing database after completion of other verifications flows. If design does not full fill DFM sign-Off criteria then it is required to do design enhancements.

DFM sign–off criteria can be for CHIP level or per CELL level. While evaluating CHIP level score for a dfm rule, all the violation in design are combined to compute aggregated DFM SCORE while for CELL level score only the violation specific to cell level is included in the calculation. DFM scoring flow also generate improvability markers for in design fixing. This will be used if design not able to full fill the DFM sign-off criteria. These DFM improvability markers are opportunistic markers which only be generated if sufficient white space is available around the violation. This helps to avoid any change in existing routing and used to do in-design fixing.

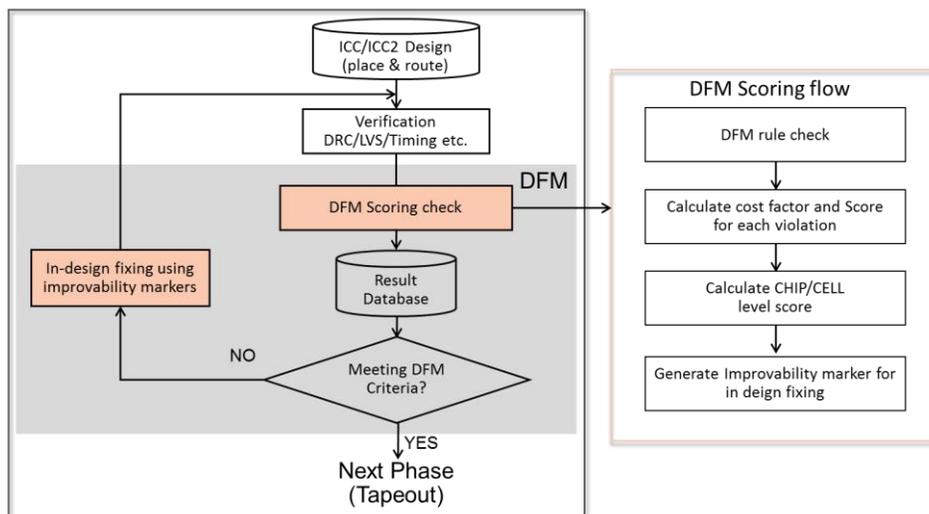

**Figure 2: DFM scoring flow integrated in design cycle.**

## 3.2 Scoring methodology

In this paper we have used an equation based scoring technique to evaluate the dfm violation severity and provide an equivalent score for it. In This the first step involves finding DFM violations which is very much same as finding DRC violation. In our example we have used M1-V1 enclosure rule termed as "RuleA". The related ICV runset shown in fig 3. This runset uses "enclose_error" command to return enclosure error markers for V1 via which are violating recommended M1 enclosure value referred as "vRuleA_RR". Fig 4 showing an example for enclosure error.

```
gRuleA = enclose_error(
sV1, aM1, < vRuleA_RR,
extension = RADIAL,
extension_look_past = POINT_TO_POINT,
intersecting = { ACUTE },
look_thru = NOT_CONTAINED
);
```

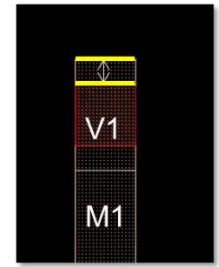

**Figure 3: Sample ICV runset for enclosure rule.**            **Figure 4: Enclosure rule violation example.**

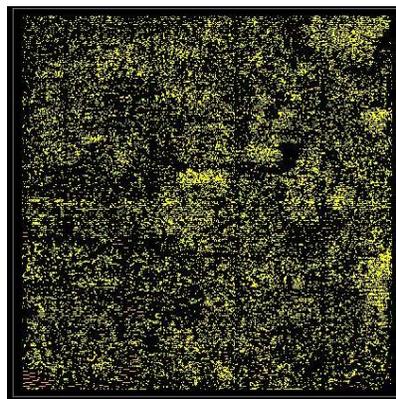

**Figure 5: DFM violation highlighted across full design.**

Fig 5 shows all the violation highlighted in full design. In this case total numbers of violations are ~38397. Fig 6 shows distribution of violation on the basis of drawn margin, as we can observe majority of violation are drawn on DRC value. This is result because most of the router constrained at DRC values for optimized usage of routing tracks and avoid congestion. In the second step these drawn margin are used to calculate cost factor for all the violations.

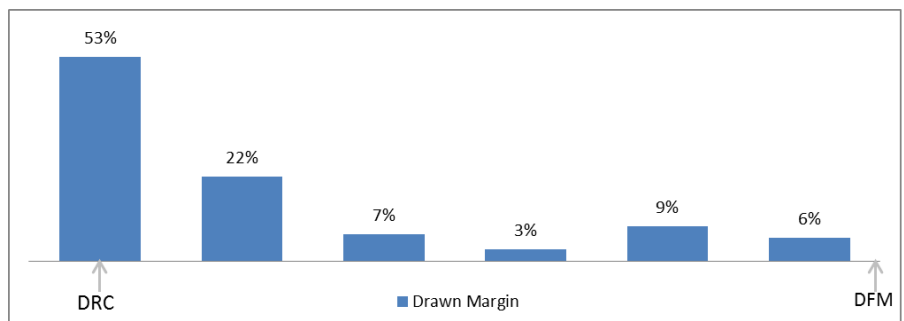

**Figure 6: Distribution of violation on the basis of drawn margin.**

The cost factor value defines the severity of DFM violation based on its sensitivity toward process margin and calculated using cost factor model. Higher cost factor represents higher severity of violation. Cost factor be evaluate as a function of Drawn value, DRC Value and DFM value, as shown in Eq1.

$$Costfactor = f(Drawn_{Value}, DRC_{Value}, DFM_{Value})$$

**Equation 1: Cost factor equation.**

Based on rule failure mechanism some rules are more sensitive towards DRC margin whereas some are bit tolerant. So it is very necessary to characterize cost factor model correctly based on rule sensitivity towards process margin.

Fig 7 shows plots comparing three hypothetical cost factor models. All three cost factor will provide different values for same drawn margin. In the figure, we can observe model CF3 has high slope towards DRC and low slope towards DFM value. This tells the rule with CF3 model is more sensitive towards DRC value and less sensitive toward DFM value. On the other hand CF1 has constant sensitivity slope across DRC to DFM values.

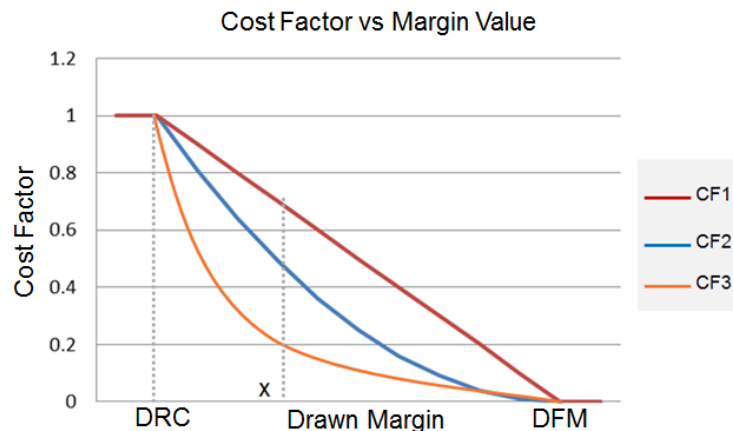

**Figure 7: Cost factor vs Margin plot.**

In Eq 1 Drawn value represents drawn enclosure value whereas DRC and DFM value represents respective required enclosure values. The cost factor applied only when Drawn value is greater the DRC value. If drawn value is smaller the DRC value than Cost factor will not applied and a default value of "1" given to that violation.

To implement cost factor calculation in ICV we have used "drc_feature_error" command as shown in fig 8. The "drc_feature_error" command uses DFM error markers as a "primary layer" and "output_form_layer" options and further process error marker using drc_function which remotely calculate cost factor value for each error marker. In this example we have used "cal_cost_fuct" as drc_function to return the cost factor values.

```
//calculate cost factor for all the violations
gRuleA_ERR = drc_features_error(
primary_layer = gRuleA,
secondary_layers = {  },
output_from_layer = gRuleA,
include_touch = NONE,
drc_function = cal_cost_func
);
```

**Figure 8: Sample ICV runset to calculate cost factor.**

SNUG 2017This function first use the df_get_current_data and df_error_layer command to retrieve error markers followed by the df_error_sum_distance command to extract the enclosure value. This enclosure value used in cost factor equation which is described in eq1.1. The cost factor equation return the cost factor value which is saved in output layer along with primary error markers gRuleA_ERR using df_save_property and df_save_data commands.

```
cal_cost_func : function (void) returning void
{primary_data = df_get_current_data();
rlayer1_set = df_error_layer(primary_data, "");
distanceL1 = df_error_sum_distance(rlayer1_set);
vCF = dblqcd ( dblgtd ( distanceL1 , vRuleA_GR ) ,
Provide here the equation to calculate cost factor;
if (double_constraint_contains(>= 0, vCF) && !(isinf(vCF) ||
isnan(vCF)))
  {   df_save_properties(primary_data, { { "CF", vCF } });
     df_save_data(primary_data);
  }
}
```

**Figure 8.1: Sample ICV runset to calculate cost factor.**

This output layer is used in next step to calculate score for each violation as shown in fig 9. For this another drc_features_error command used which parse error layer generated from previous step as "primary layer" and "output_from_layer" along with new drc_function. In this example we have used cal_score_func to remotely calculate score based on cost factor property for each layer. The function extracts the saved cost factor property and calculate score value. The score value is inverse of cost factor value. If violation has higher cost factor value then the associated score for that violation is lower. The function also extract the drawn enclosure value and save this as Drawn_Value property along with Score value property using df_save_properties and df_save data commands.

```
rRuleA_ERR @= { @ "RuleA_err";
drc_features_error(
primary_layer = gRuleA_ERR,
secondary_layers = {  },
output_from_layer = gRuleA_ERR,
include_touch = NONE,
drc_function = cal_score_func );
}
// calculate score for each violation
cal_score_func : function (void) returning void
{
primary_data = df_get_current_data();
layer1_set = df_error_layer(primary_data, "");
distanceL1 = df_error_sum_distance(layer1_set);
vDRAWN_VALUE = distanceL1;
df_report_double(primary_data, name = "Drawn_Value" , value =
vDRAWN_VALUE );
sum_propL1_CF_valid = df_get_error_sum_double_property(layer1_set,
"CF", sum_propL1_CF);
vSCORE = 1 - sum_propL1_CF;
df_report_double(primary_data, name = "Score" , value = vSCORE );
df_save_data(primary_data);
}
```

**Figure 9: Sample ICV runset to calculate score value for each violation.**

*Page 8*                                                                    *In design DFM rule scoring and fixing method using ICV*

Fig 10 shows snap shot of layout error summary file showing list of all the violations along with saved properties of Drawn_Value and Score value.

```
-------------------------------------------------------------------------------
RuleA_err
-------------------------------------------------------------------------------

- - - - - - - - - - - - - - - - - - - - - - - - - - - - - - - - - - - - - - -
Structure          ( lower left x, y )   ( upper right x, y )
                                                              Report    = Value
- - - - - - - - - - - - - - - - - - - - - - - - - - - - - - - - - - - - - - -
SEL_FSDPMQ_D_2     (0.4206, 0.2100)      (0.5294, 0.2450)
                                                              Drawn_Value = ▇▇▇▇▇
                                                              Score       = 0.5000
SEL_FSDPMQ_D_2     (0.4135, 0.3150)      (0.4965, 0.3450)
                                                              Drawn_Value = ▇▇▇▇▇
                                                              Score       = 0.0000
SEL_FSDPMQ_D_2     (0.2504, 0.5050)      (0.3496, 0.5250)
                                                              Drawn_Value = ▇▇▇▇▇
                                                              Score       = 0.0000
```

**Figure 10: Sample result summary for each rule with score property annotated.**

Next step involves calculation of total DFM_SCORE using scoring model. There are multiple models which can be used for DFM scoring. In our approach, we have used Poisson yield equation [4]. This model based on Poisson probability distribution function and used for calculate the wafer yield score. The Poisson model requires the defects are uniformly and randomly distributed, a model which tends to predict yields [4].

$$DFM\_Score_a = \prod_{i=1}^{n} f(Score_i, FR_a)$$

**Equation 2: DFM Scoring equation.**

Eq 2 showing DFM scoring equation which is used in our example. In this equation "n" represent the total number of violations, "FR" represents Failure Rate. Based on this equation all the violations induvial scores are put together and multiplied by the Failure rate. FR can be being understood as hit rate. For example, assume there are 1e9 feature count exist in the die and there is only 1 feature failed at silicon so the failure rate becomes 1 per part of 1e9 (1e-9). Failure Rates can be different for different rules. During process development cycle FR also improves with process maturity.

To implement chip level DFM_SCORE in ICV, we use "dfm_feature" command as shown in fig 11. This command used DFM violation as primary layer with context variable set to "TOP". This help to accumulate all the violation from top to bottom level by resolving all the hierarchy. dfm_fetaure command call another dfm function and parse the DFM violation as primary layer to return DFM_SCORE and save the results in output layer. This dfm_function uses dfm_count command to get the count of dfm violations and if count is greater than "0" then it calculate DFM SCORE using the Poisson yield equation described previously. To get the sum of all cost factor property dfm_get_sum_double_property command is used and to get design area dfm_window_area command is used. In the given example we use 0.99 as sign-off value and if DFM_SCORE is less than 0.99 the the rule is reported in the result data base.

```
//Scoring function
dfm_scoring_func : function (void) returning void
{
DFM_SCORE = ((dfm_count("primary_layer") > 0) ?
Provide DFM scoring equation Here;
if (DFM_SCORE < 0.99) { @ "Status: FAIL";
dfm_report_double( name = "DFM_SCORE" , value = DFM_SCORE );
dfm_save_data();
}
}
//calcultae DFM SCORE for CHIP level
rRuleA_CHIP_Score @= { @ "RuleA_CHIP_Score";
dfm_features(
layer_ids = {"primary_layer" => gRuleA_ERR},
dfm_function = dfm_scoring_func,
context =  TOP
);
}
```

**Figure 11: Sample ICV runset for CHIP level DFM SCORE calculation.**

Fig 12 shows the snapshot of result data base. In this we can observe the structure column showing TOP cell name followed by total x, y coordinates of full design. Report column shows DFM_SCORE value. In this example chip level score is 0.9692 which is smaller than sign-off value and hence reported in the design.

```
--------------------------------------------------------------------------
RuleA_CHIP_Score
--------------------------------------------------------------------------

- - - - - - - - - - - - - - - - - - - - - - - - - - - - - - - - - - - - -
Structure   ( lower left x, y ) ( upper right x, y )
                                                    Report    = Value
- - - - - - - - - - - - - - - - - - - - - - - - - - - - - - - - - - - - -
or1200_top (5.0000, 0.3450)    (258.4650, 258.2950)
                                                    DFM_SCORE = 0.9692
```

**Figure 12: Sample result summary for DFM SCORE for chip level.**

This methodology is also being applied to calculate the cell level scoring.  To compute cell level score another dfm_fetaure command is used as shown in fig 13. The mechanism is very similar as chip level scoring except in the context is set to BY_CELL. This setting allow ICV tool to compute violation accumulated based on cell individual placements without resolving any hierarchy. By doing cell level score analysis we can get the list of cell which are failing the sign-off criteria. This help designer to filter out cells with low DFM_SCORE

```
//calcultae DFM SCORE for CELL level
rRuleA_CELL_Score @= { @ "RuleA_CELL_Score";
dfm_features( layer_ids = {"primary_layer" => gRuleA_ERR},
dfm_function = dfm_scoring_func, context = BY_CELL);
}
```

**Figure 13: Sample ICV runset for CELL level DFM SCORE calculation.**

Fig 14 shows the snapshot of CELL level summary result. Likewise CHIP level score we can see CELL level DFM SCORE for different cell which failing the criteria.

```
--------------------------------------------------------------------------
RuleA_CELL_Score
--------------------------------------------------------------------------

- - - - - - - - - - - - - - - - - - - - - - - - - - - - - - - - - - - - -
Structure         ( lower left x, y ) ( upper right x, y )
                                                           Report    = Value
- - - - - - - - - - - - - - - - - - - - - - - - - - - - - - - - - - - - -
SEH_EN2_1         (0.0000, -0.0350)   (1.5400, 1.2950)
                                                           DFM_SCORE = 0.9523
SEL_FSDPMQ_D_2    (0.0000, -0.0350)   (4.7600, 1.2950)
                                                           DFM_SCORE = 0.8607
SEH_EN2_S_1       (0.0000, -0.0350)   (1.5400, 1.2950)
                                                           DFM_SCORE = 0.9523
SEH_EO2_G_2       (0.0000, -0.0350)   (1.8200, 1.2950)
                                                           DFM_SCORE = 0.9595
SEL_FSDPQO_D_2    (0.0000, -0.0350)   (3.7800, 1.2950)
                                                           DFM_SCORE = 0.9706
```

Figure 14: Sample result summary for CELL level DFM SCORE.

### 3.3 Executing DFM scoring check using command line option

Likewise DRC run, users can also run DFM scoring checks through command line using signoff_drc command and setup signoff_options using set_physical_signoff_options command

icc_shell>   set_physical_signoff_options   -exec_cmd   icv   -drc_runset DFMScoringKit.rs

icc_shell> signoff_drc

DFM_Runset.rs is referring to ICV run set having all the previously described commands and functions.

### 3.4 Executing DFM scoring check using VUE GUI

DFM scoring checks can also be triggered using VUE gui as shown in fig 15. To launch first need to load the design and go to verification and then click on IC Validator VUE, this will bring VUE GUI then provide DFM Runset and Run directory and click on "Execute" button to launch the run.

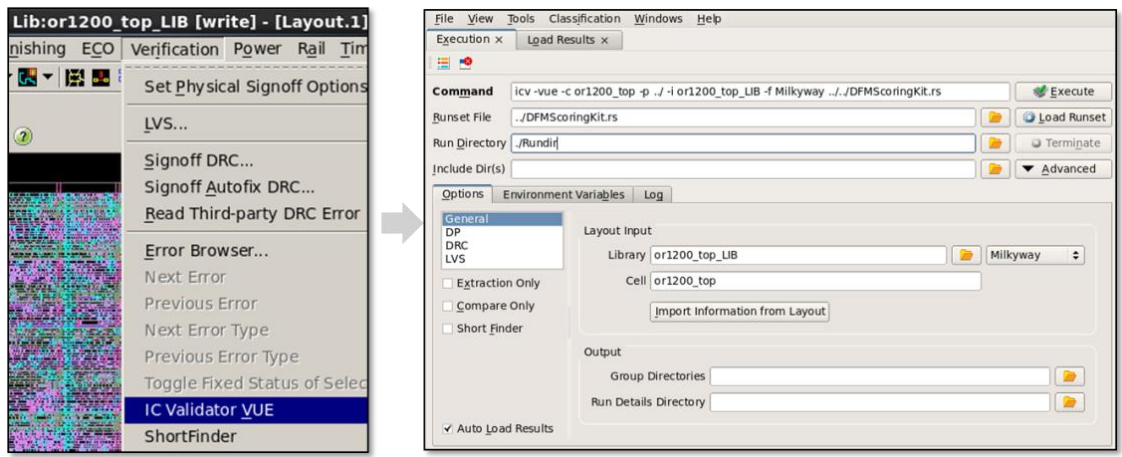

Figure 15: Launching DFM scoring kit using VUI gui.

Violation browser is a very handy and useful tool. User can filter rules and and sort based on DFM SCORE. User can navigate violations and review accordingly. Fig 16 shows chip level DFM_SCORE for RuleA which is 0.9692. This can indicate designers the enclosure rule may have a lot of violations and does not meet sign-off criteria (assume sign-off criteria >= 0.99)

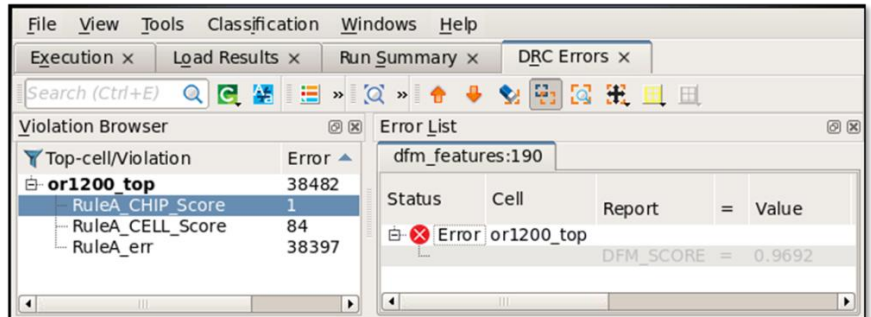

Figure 16: Reviewing CHIP level score using VUE gui.

Designers can navigate all the violation further to understand and sort them based on score value as shown in fig 17.

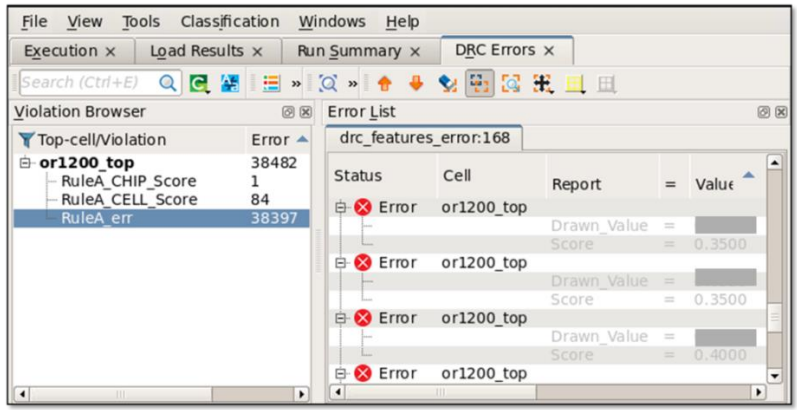

Figure 17: Reviewing all violations using VUE gui.

Similarly CELL level results and can also be used to analyze cell which are failing the sign-off criteria and sort based on low to high score as shown in fig 18. In this example, we can observe that cell with name "SEN_FSDPMQ_D_1" has lowest DFM_SCORE 0.8568 among the entire cell.  And if this cell is critical then user might need to review the violations inside this cell.

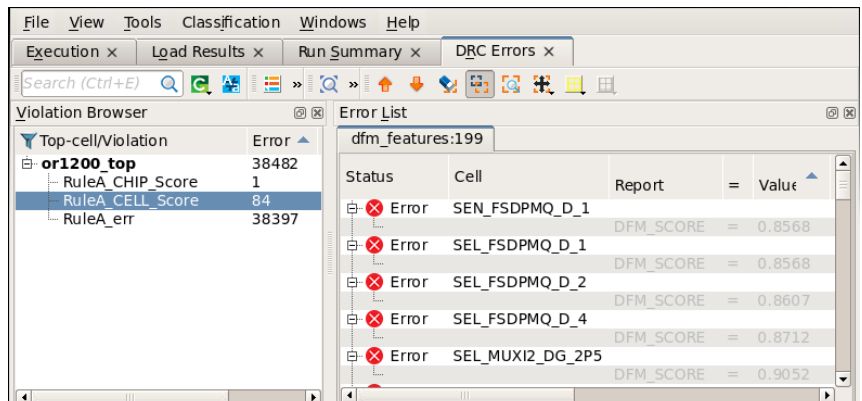

Figure 18: Reviewing CELL level score using VUE gui.

## 3.5 In-design DFM fixing technique

In contrast with checking DFM violation it is also be interesting to understand how to fix the DFM violations. This help designer to improve their design and improve DFM SCORE. In this paper we have use Improvability approach for fixing technique. This approach uses the existing routing and available white space near each dfm violations. If sufficient space is available then feature margin will be increased. In metal to via enclosure example our interest is to know if enclosure margin can be increased. To evaluate this we use ICV drc commands as shown in fig 19. In this method first all the edges which need to grow are derived based on dfm violations using `enclose_edge` command. Fig 20 shows the edge marker for enclosure violation.

```
gRuleA_err_edge1 = enclose_edge(
sV1, aM1, < vRuleA_RR,
extension = RADIAL,
extension_look_past = POINT_TO_POINT,
intersecting = { ACUTE },
look_thru = NOT_CONTAINED,
output_layer = LAYER2
);
```

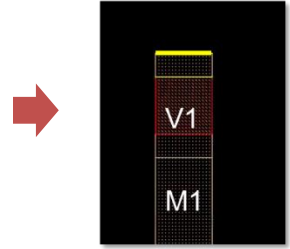

**Figure 19: Sample ICV runset to get edge layer for improvement.**  **Figure 20: Error edge to be enhanced.**

Next step involves the calculation of how much more enclosure is needed to full fill DFM enclosure value. There can be multiple ways to implement this. In our approach we first identify all the M1 edges which are touching these error edges using touching_edge command. Then we have used drc_features_edge command and parse M1 edges as primary layer along with DFM error marker derived previously. This command calls drc_function to calculate required enclosure value to full fill DFM requirement and save this value as property in the output layer. This output layer is further used by another ICV command which is edge_size_by_property. This command is very usfull and can size an edge based on save property in it. In this case all the M1 edges sized outside by property stored in previous step as shown in fig 21. This give enhaced M1 shapes as improbaility marker.

```
gRuleA_m1_edge = aM1 touching_edge gRuleA_err_edge1;
get_imp_marker : function (void) returning void
{primary_data = df_get_current_data();
rlayer2_set = df_error_layer(primary_data, "layer2");
distanceL2 = df_error_sum_distance(rlayer2_set);
vReq_Enc_Value = ( vRuleA_RR - distanceL2 );
df_save_properties(primary_data,{{"Req_Enc_Value", Req_Enc_Value} });
df_save_data(primary_data);
}
gRuleA_impMarker = drc_features_edge(
primary_layer = gRuleA_m1_edge,
secondary_layers = { "layer2" => gRuleA_ERR },
output_from_layer = gRuleA_m1_edge,
include_touch = EDGE,
drc_function = get_imp_marker
);
rRuleA_ImpMarker1 = edge_size_by_property(
gRuleA_impMarker,
outside_property = "Req_Enc_Value"
);
```

**Figure 21; Sample ICV runset to size edge layers as improvability marker.**

These shapes are needed to be filtered as they may create other DRC or DFM violations. So to achieve this we can use multiple filtering rule check. In our example we have used only spacing check. If the enhanced shapes are creating any spacing violation then those shapes are considered as invalid and been removed as shown in fig 22.

```
//Filter Invalid Imp Markers
invalid_ImpMarker = external2(
rRuleA_ImpMarker1, aM1, < vM1_Spacing, extension = RADIAL,
intersecting = { ACUTE },relational = { POINT_TOUCH }
);
```

**Figure 22: Sample ICV runset to filter invalid improvability marker.**

After filtering, all the markers are stored in output layer in result database as shown in fig 23. Fig 24 shows the example of improvability marker generated.

```
//Write fixing marker in to the database
rRuleA_Imp_ERR @= { @ "RuleA_IMP_Markers";
not_interacting (
rRuleA_ImpMarker, invalid_ImpMarker
);
}
```

**Figure 23: Sample ICV runset to write improvability markers in result database.**

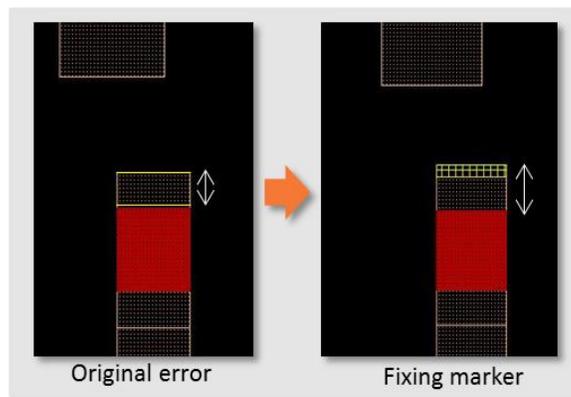

**Figure 24: Showing example of improvability marker.**

These improvability marker can also be useful to calculate Improvability score by combining it with orignal shapes and perform DFM scoring. Fig 25 shows the flow overview for Improvability Scoring

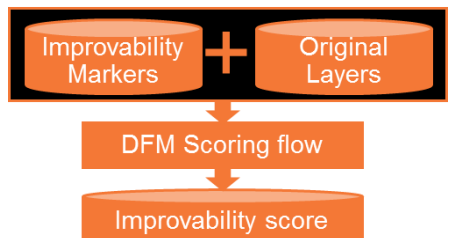

**Figure 25: Improvability Scoring Flow.**

To merge the improvability marker user can simply use "OR" operation as shown in fig 26.

```
//Merge IMP markers with ORG geometry
New M1 = aM1 or rRuleA ImpMarker;
```

**Figure 26: Sample ICV runset to merge improvability markers with original geometries.**

Performing DFM scoring based on new merged layer can produce Improvability score as shown fig 27. In this case we can observe the improbility scope is very high and new score jumps to 0.9947. Hense it will be good idea to implement these markers in design.

```
--------------------------------------------------------------------
RuleA_CHIP_Score_imp
--------------------------------------------------------------------

- - - - - - - - - - - - - - - - - - - - - - - - - - - - - - - - - -
Structure  ( lower left x, y ) ( upper right x, y )
                                                 Report     = Value
- - - - - - - - - - - - - - - - - - - - - - - - - - - - - - - - - -
or1200_top (5.0000, 0.3450)    (258.4650, 258.2950)
                                              DFM_SCORE = 0.9947
```

**Figure 27: Improvability score output.**

To implement improvability markers in design we have used ICC script. This script extract and store imp marklers in command file. The script first trigger DFM scoring check and extract all the error markers along with x,y coordinates and net name and generate another file having "create_net_shape" command to write imp markers in design".

```
set_physical_signoff_options -exec_cmd icv -drc_runset
DFMScoringKit.rs
signoff_drc -run_dir RunDir1 -error_view DFM_SCORE.err
set view [gui_open_error_view -name DFM_SCORE.err  ]
set fp [open "write_imp_shapes" w]
set errors [get_drc_errors -type RuleA_IMP_Markers -error_view $view -
quiet]
 foreach_in_collection id $errors {
 set bbox [get_attribute $id bbox];
 set nets [get_attribute -class drc_error $id nets]
   foreach_in_collection net_id $nets {
    set net_name [get_attribute $net_id name]
   }
puts $fp "create_net_shape -type wire -bbox $bbox -layer M1 -net
$net_name";
}
close $fp
Icc_shell> source extract_imp_marker.tcl
```

The generated file should look like below. This file create imp shapes in design when sourced

```
create_net_shape  -type  wire  -bbox {{193.811  154.975} {193.843
154.985}} -layer M1 -net net42054
create_net_shape  -type  wire  -bbox {{193.811  154.903} {193.843
154.913}} -layer M1 -net net56289
………………………………………
```

```
Icc_shell> source write_imp_shapes
```

After implementing the improvability marker back in design it is of interest to compare improvement in DFM SCORE. Fig 25 shows the comparison of DFM results between original and enhanced design. We can see the DFM violation reduced after implementation of Improvability markers and DFM SCORE is also increased to 0.9947 which passes the DFM criteria.

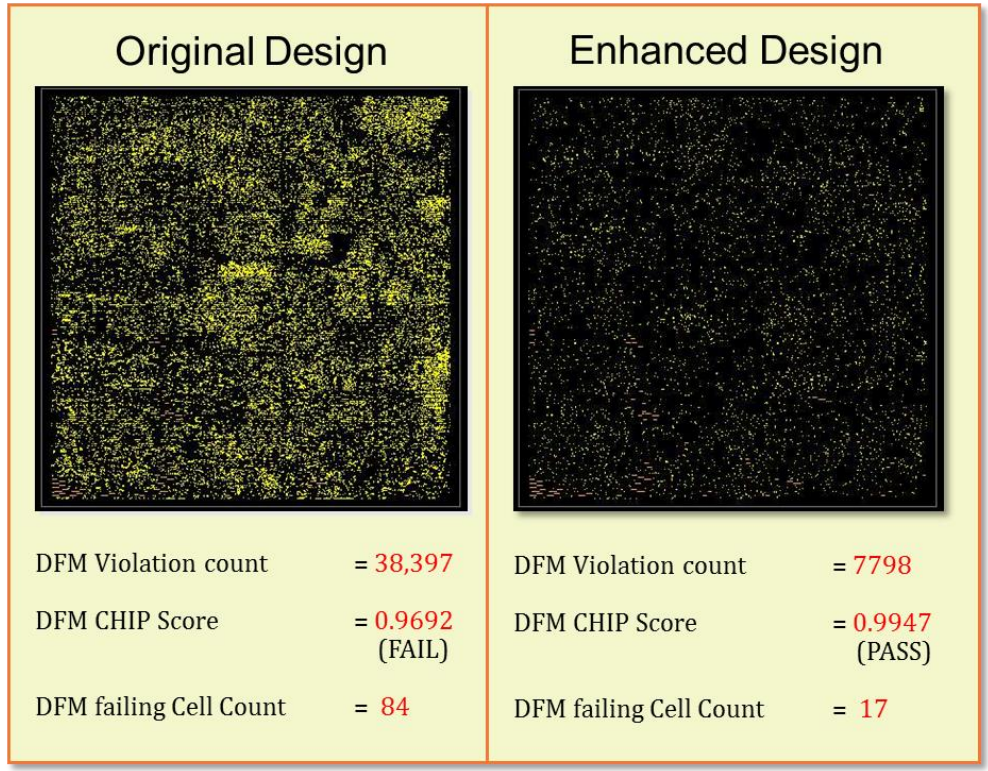

Figure 28: Showing comparison between original design vs Enhanced design.

## 4. Conclusions

The presented DFM scoring approach is essential in terms of understanding design quality. Doing such DFM scoring outside design environment will created multiple iteration in closing the design loop and delay in overall time-to-tapeout. This also creates a dependency on 3rd party EDA tools to do such DFM scoring checks. The demonstrated in-design DFM scoring and fixing approach resolves these issues and can easily be adapted early in the design cycle. By using this approach designs can address DFM requirements and able to develop DFM compliance designs.

## 5. References

[1] Using in-design physical verification to reduce tapeout schedules - Edadesignline-2010.

[2] Synopsys, IC Validator Reference Manual Version M-2016.12-SP2, March 2017.

[3] Piyush Pathak, Sriram Madhavan, Shobhit Malik, Lynn T. Wang, Luigi Capodieci.,"Framework for identifying recommended rules and DFM scoring model to improve manufacturability of sub-20nm layout design" SPIE 8327, Design for Manufacturability through Design-Process Integration VI, 83270U (15 March 2012)

[4] Cunningham, J. A. "The use and evaluation of yield models in integrated circuit manufacturing,"Semiconductor Manufacturing IEEE Transactions 3(2), 60-71 (1990)